\documentclass[runningheads,a4paper]{llncs}

\usepackage{amssymb}
\usepackage{amsmath}
\usepackage{graphicx}

\usepackage{epsfig}
\usepackage{graphicx}% Include figure files
\usepackage{dcolumn}% Align table columns on decimal point
\usepackage{bm}% bold math
\usepackage{tikz}
\usepackage[all]{xy}
\usetikzlibrary{arrows,shapes}
\usepackage{url}
\usepackage{centernot}
\usepackage{stmaryrd}

\begin{document}

\mainmatter  % start of an individual contribution

\title{Fast multipole networks
}

\titlerunning{Fast multipole networks}

\author{
Steve Huntsman\inst{1}
}
\authorrunning{S. Huntsman}

\institute{
BAE Systems FAST Labs\\
\email{steve.huntsman@baesystems.com}
}


\toctitle{Fast multipole networks}
\tocauthor{S. Huntsman}
\maketitle

\begin{abstract}
Two prerequisites for robotic multiagent systems are mobility and communication. \emph{Fast multipole networks} (FMNs) enable both ends within a unified framework. FMNs can be organized very efficiently in a distributed way from local information and are ideally suited for motion planning using artificial potentials. We compare FMNs to conventional communication topologies, and find that FMNs offer competitive communication performance (including higher network efficiency per edge at marginal energy cost) in addition to advantages for mobility.
%\keywords{fast multipole method, network, random geometric graph, Delaunay triangulation, multirobot system, swarming}
\end{abstract}

\section{\label{sec:Introduction}Introduction}

A multirobot system \cite{Knorn2016} is a group of autonomous, networked robots. In order to achieve a complex goal such as swarming \cite{Chung2018}, the system requires distributed coordination of both mobility and communication, among other objectives. This is nontrivial, and ``[e]fficient networking of many-robot systems is considered one of the grand challenges of robotics'' \cite{Minelli2019}. The respective enabling technologies for mobility and communication are path planning and mobile \emph{ad hoc} networks (MANETs) \cite{Loo2016}. While networks are inevitably analyzed from the perspective of graph theory \cite{Barrat2008}, path planning may be considered in either graph-theoretical \cite{Mesbahi2010} or continuous settings. Meanwhile, because geometrical considerations such as distance and motion strongly influence the structure of MANETs, it is natural to try to address mobility and communication for multirobot systems together, e.g., as in \cite{Stephan2017}. Much effort has focused on connectivity maintenance in situations where, e.g. multirobot systems maintain periodic connectivity \cite{Hollinger2010} or communicate by physically meeting \cite{Kantaros2019} while pursuing a motion objective, or maintain continuous connectivity relative to a fixed set of access points \cite{Ghaffarkhah2011,Kantaros2016a,Kantaros2016b}. Additionally, co-optimization of communication and motion or coverage for an individual robot have been considered in \cite{Yan2011,Minelli2019}. More recently, tree-based approaches for connectivity maintenance have been considered in \cite{Krupke2015,Majcherczyk2018,Varadharajan2019}.

In this paper, we \emph{assume} connectivity is possible (by using more energy if necessary) without any optimization, and we introduce a class of network backbones that can be trivially formed using an efficient local motion planning technique. These \emph{fast multipole networks} (FMNs) to support both mobility and communication within a unified framework. The basic idea is to follow common practice in modeling robots, goals, and obstacles as (superpositions of) charged particles satisfying the Laplace equation $\nabla^2 \phi = 0$ \cite{Connolly1990,Kim1992,Pimenta2005} and exploit the \emph{fast multipole method} (FMM), an efficient algorithm for simulating particle dynamics \cite{Beatson1997,Board2000,Greengard1987}, to simultaneously determine a sparse network topology that supports efficient communication. The animating principle that the far-field behavior of point charges \cite{Jackson1998} should determine a communication topology is geometrically natural. More surprisingly, we shall demonstrate that it leads to network topologies that perform well in their own right, with higher network efficiency per edge (at marginal energy cost) than standard topologies that ignore mobility.

After briefly reviewing the artificial potential approach to path planning in \S \ref{sec:Potentials} and the FMM in \S \ref{sec:FMM}, we introduce FMNs in \S \ref{sec:FMN}, and compare them to conventional MANET topologies in \S \ref{sec:Evaluation} before making concluding remarks in \S \ref{sec:Remarks}.

\section{\label{sec:Potentials}Artificial potentials}

The use of artificial potentials in motion and path planning has a long history, most frequently identified as beginning with \cite{Khatib1985}. The basic idea is to design a potential $\phi$ such that the equation of motion $m \ddot x = -\nabla \phi$ results in a desired trajectory $x$. Towards this end, goals and obstacles are respectively modeled by attractive and repulsive terms contributing to the total potential $\phi$. Depending on circumstances, we may choose to model the robots as ``sources'' with potentials of their own (e.g., to avoid collisions), or as passive ``targets'' that simply move along the gradient of an ambient potential.

In general, we might consider essentially arbitrary forms for each term to produce very detailed behavior. Alternatively, we might rely on a single simple form for all the terms. Our approach is in the latter vein. The relative strengths and spatial distribution of these terms are chosen to establish priorities, spatially extended features, etc. In order to represent sufficiently complex spatial relationships along these lines, it is helpful to have an algorithmic framework that scales better in total computational effort, parallelism, and locality than evaluating $O(N^2)$ interactions, since the number $N$ of terms in the potential can be much larger than the number of robots involved.

Besides these computational concerns, a problem with using artificial potentials that was identified at an early stage is the possible presence of local minima in the potential field that can trap agents \cite{Koren1991}. To remedy this by construction, the notion of a \emph{navigation function} that has a single minimum at the goal was developed, along with algorithms for constructing such functions \cite{Rimon1992}. A particularly simple way to avoid local minima while using a single form for all the potential terms is with a superposition of harmonic potentials \cite{Connolly1990,Kim1992,Pimenta2005}, i.e., solutions to the Laplace equation $\nabla^2 \phi = 0$, with a dominant term at the goal. 

This is most readily achieved through a discrete (if perhaps quasi-continuous) superposition of \emph{point charges}, i.e. potentials of the form $-qV(|x-x_0|)$ (the sign is for physical reasons), where the \emph{fundamental solution} $V(|x|)$ to the Laplace equation is defined by $\nabla^2 V(|x|) = \delta(x)$, and as usual $\delta$ indicates the Dirac delta distribution \cite{Taylor1996}. For $\mathbb{R}^d$, it turns out that $V'(r) = 1/A_d(r)$, where $A_d(r)$ is the Minkowski content (i.e., generalized perimeter, surface area, etc.) of the sphere of radius $r$ in $\mathbb{R}^d$. Choosing the most convenient constants of integration, for $d = 2$ we have $V(r) = \frac{1}{2\pi} \log r$, and for $d = 3$ we have $V(r) = -1/4\pi r$.

\section{\label{sec:FMM}The fast multipole method}

\begin{figure}[htbp]
\begin{center}
\includegraphics[trim = 60mm 95mm 60mm 92mm, clip, width=.45\textwidth,keepaspectratio]{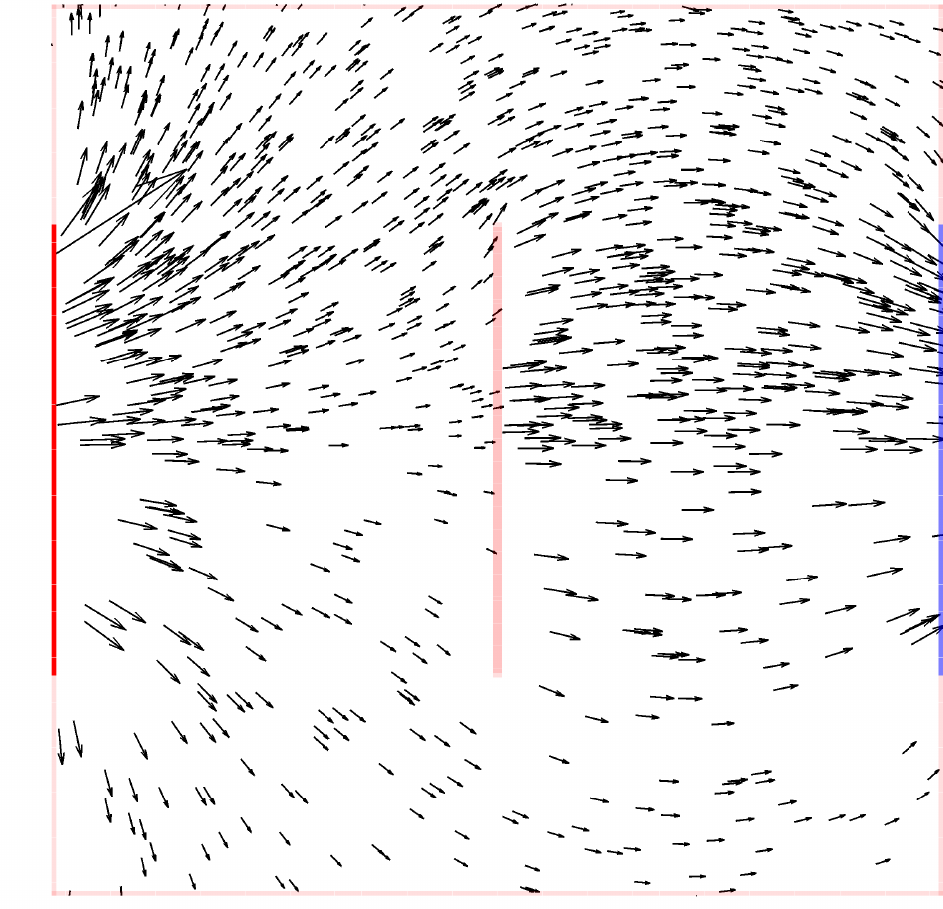}% Here is how to import pix
\includegraphics[trim = 50mm 95mm 50mm 85mm, clip, width=.54\columnwidth,keepaspectratio]{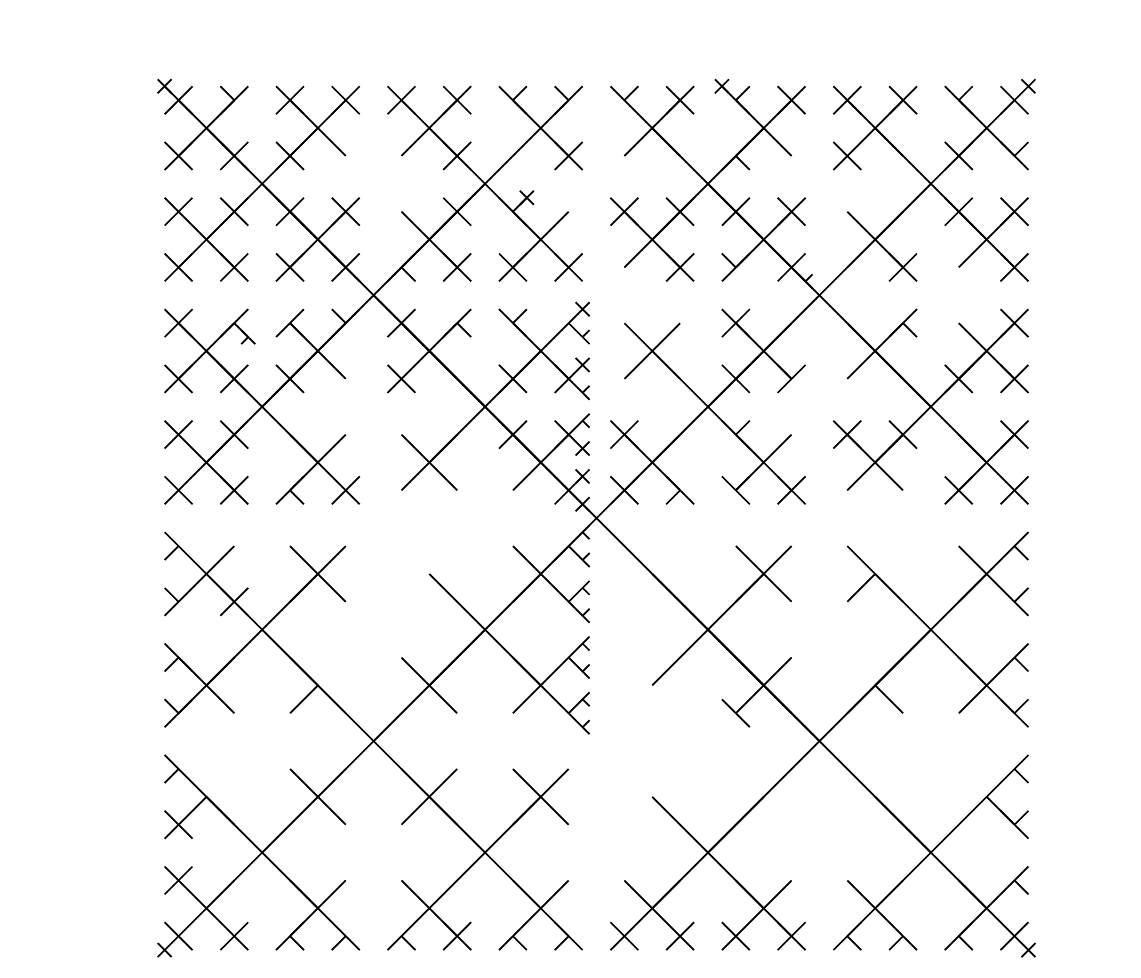}% Here is how to import pix
\caption{ \label{fig:FMM} (L) A toy scenario in $[-1,1]^2$. Goals are modeled by negative charges and shown in {\color{blue}blue}; obstacles are modeled by positive charges and shown in {\color{red}red}. Opacity indicates relative magnitude. $10^3$ robots are modeled by test points (versus, e.g., test charges of small positive sign) and their locations and velocities indicated by black gradient vectors of the artificial potential. The target locations are distributed as $\frac{4}{5}U(\text{top half}) + \frac{1}{5}U(\text{bottom half})$, where here $U$ indicates a uniform distribution. (R) The quad-tree associated to the scenario on the right. Varying the desired precision in the FMM has very little effect on this tree; as a practical matter it can be assumed unique. %(Bottom right) The associated spatial discretization, with relative number of test points indicated. Note that regions without test points do not have ``leaf boxes.''
} 
\end{center}
\end{figure}

Naive simulation of $N$ interacting point charges (e.g., the goals and obstacles modeled in Figure \ref{fig:FMM}) requires computing the interactions of each pair of charges, and hence $O(N^2)$ operations per time step, which is prohibitive for large-scale $N$-body simulations. The FMM \cite{Beatson1997,Board2000,Greengard1987} enables the simulation cost to be reduced to $O(N)$ with an extremely high degree of locality and parallelism \cite{Greengard1990}. 
\footnote{
For the calculations in this paper, we used the very user-friendly library FMMLIB2D, available at \url{https://cims.nyu.edu/cmcl/fmm2dlib/fmm2dlib.html}.
}

The key ideas underlying the FMM are
\begin{itemize}
\item[i)] a specification of accuracy (for truncating expansions in a controlled way);
\item[ii)] decomposing space hierarchically to get \emph{well-separated} charge clusters;
\footnote{
Two clusters of points $\{x_j\}$ and $\{y_k\}$ are well-separated iff there exist $x_0, y_0$ such that $\{x_j\} \subset B^\circ_{x_0}(r)$ and $\{y_k\} \subset B^\circ_{y_0}(r)$ with $|x_0 - y_0| > 3r$: here $^\circ$ denotes interior. Two squares with side length $r$ are well-separated iff they are at distance $\ge r$.
}
\item[iii)] representing well-separated clusters of point charges with multipole expansions that maintain a desired approximation error $\varepsilon$ with as few ($\lceil \log_2(1/\varepsilon) \rceil$) terms as possible, leaving nearby particles to interact directly.
\end{itemize}
In particular, the FMM recursively builds a quad-tree (Figure \ref{fig:FMM}; in three dimensions, an octo-tree is used instead) whose leaves are associated with boxes and truncated multipole expansions. This tree approximates a (typically much) finer tree whose leaves are associated with individual point charges that are well-separated and their monopoles. Importantly, the FMM tree topology essentially ignores the values of charges, depending only on the desired level of accuracy $\varepsilon$ 
\footnote{
Though in principle the desired level of accuracy can be affected by charge values, this situation is sufficiently pathological that we can safely disregard it in practice.
}
and the locations of the charges. 

The computationally expedient part of the FMM is to manipulate the origins and coefficients of controlled series approximations to far-field potentials for clusters of point charges that are well-separated.
More general incarnations of the FMM (see, e.g., \cite{Letourneau2014,Ying2004,Ying2006}) amount to a very efficient scheme for computing sums of the form $\sum_{j=1}^N K(x_i,\xi_j) \psi(\xi_j)$ for a given \emph{kernel} $K$: i.e., the FMM and its generalizations are essentially specialized matrix multiplication algorithms. From this perspective, item iii) in the list above separates into \cite{Beatson1997}
\begin{itemize}
\item a far-field expansion of the kernel $K(x,\xi)$ that decouples the influence of the evaluation/target point $x$ and the source point $\xi$;
\item (optionally) a conversion of far-field expansions into local ones.
\end{itemize}

The FMM's remarkable scaling performance has enabled petascale simulations of turbulence \cite{Yokota2013}, molecular dynamics \cite{Ohno2014}, and cosmological dynamics \cite{Potter2017}, and will also enable future exascale simulations across hundreds of thousands of nodes \cite{Yokota2012}. This performance makes the FMM a natural choice for large scale path planning using artificial potentials. 

Equally important for the considerations of this paper, however, are the hierarchical and spatial locality properties that the FMM exploits in order to communicate internally. The FMM's patterning of a logical intra-algorithm communication network after the spatial distribution of particles suggests that it can be used not only for large-scale multirobot path planning in complex geometries, but also to help organize the communications between robots in a distributed way. Furthermore, although the FMM's hierarchical properties might seem to imply centralization, the computational load is small enough that these functions can be easily duplicated among robots with low overhead, i.e., the FMM tree does not impose centralization.

\section{\label{sec:FMN}Fast multipole networks}

\begin{figure}[htbp]
\begin{center}
\includegraphics[trim = 50mm 95mm 50mm 85mm, clip, width=.8\textwidth,keepaspectratio]{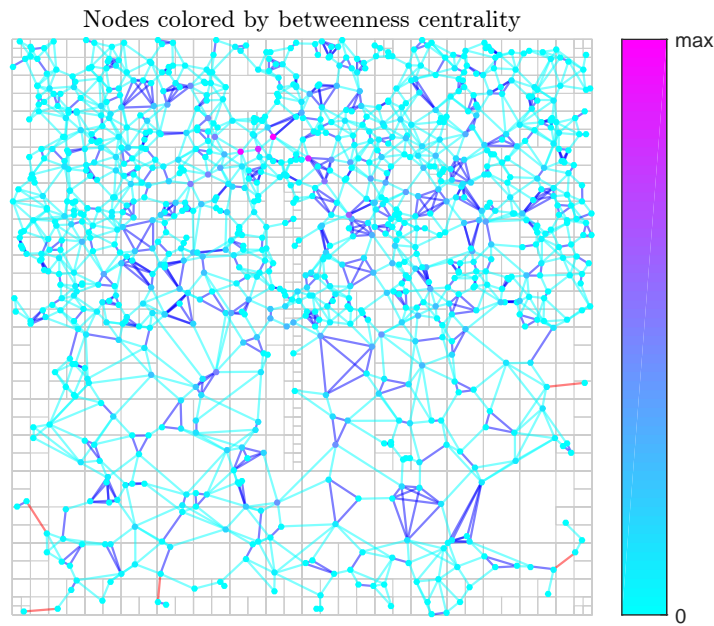}% Here is how to import pix
\caption{ \label{fig:FMNet} The FMN corresponding to the scenario in Figure \ref{fig:FMM}. Nodes are colored by betweenness centrality according to the colorbar on the right. {\color{gray}The spatial decomposition from Figure \ref{fig:FMM} is shown in gray} for reference. {\color{blue}Edges within a FMM box are blue}, while {\color{cyan}edges connecting nearest nodes in adjacent boxes are cyan} and {\color{red}edges connecting otherwise isolated nodes to their nearest neigbors are red}.} 
\end{center}
\end{figure}

We construct the \emph{fast multipole network} $FMN(\xi)$ corresponding to a configuration of points $\xi_j \in \mathbb{R}^2$ as follows: vertices correspond to the charge locations and we introduce edges that
\begin{itemize}
	\item connect all vertices in the same FMM leaf box;
	\item connect nearest vertices in adjacent leaf boxes;
	\item connect otherwise isolated vertices to their nearest neighbors.
\end{itemize}
These edges are respectively colored {\color{blue}blue}, {\color{cyan}cyan}, and {\color{red}red} in Figure \ref{fig:FMNet}.
\footnote{
The key difference between FMNs and the networks considered in \cite{Zitin2015} is that the latter are formed by inserting and permanently linking nearby charges, then dynamically evolving to obtain small-world features, whereas FMNs are (re)formed by linking nearby charges in a way that partially anticipates the next timestep of dynamical evolution. However, both types of networks exhibit aspects of small-world behavior (see \S \ref{sec:Evaluation} and \cite{Latora2001}).
}	

By construction, $FMN(\xi)$ is connected, and the information required to generate it is automatically produced by the FMM. We note that while $FMN(\xi)$ is constructed using the quad- or octo-tree of the FMM, it is very far from a tree. Rather, the FMM tree and its corresponding coarse-graining of space determines which nodes are \emph{permitted} to communicate directly.
\footnote{
Limiting permission for direct communication in FMNs can be enforced by, e.g., cognitive radios \cite{Yu2011} whose spectrum allocation cooperates with the FMM tree.
}
Within a clique of permitted communications corresponding to a leaf of the FMM tree, we may further restrict communications to avoid quadratic bandwidth overhead and/or energy, though we do not consider such tactics further here. 

\section{\label{sec:Evaluation}Evaluation}

We now introduce several families of graphs for evaluation purposes. 

\begin{figure}[htbp]
\begin{center}
\includegraphics[trim = 50mm 95mm 50mm 85mm, clip, width=.5\textwidth,keepaspectratio]{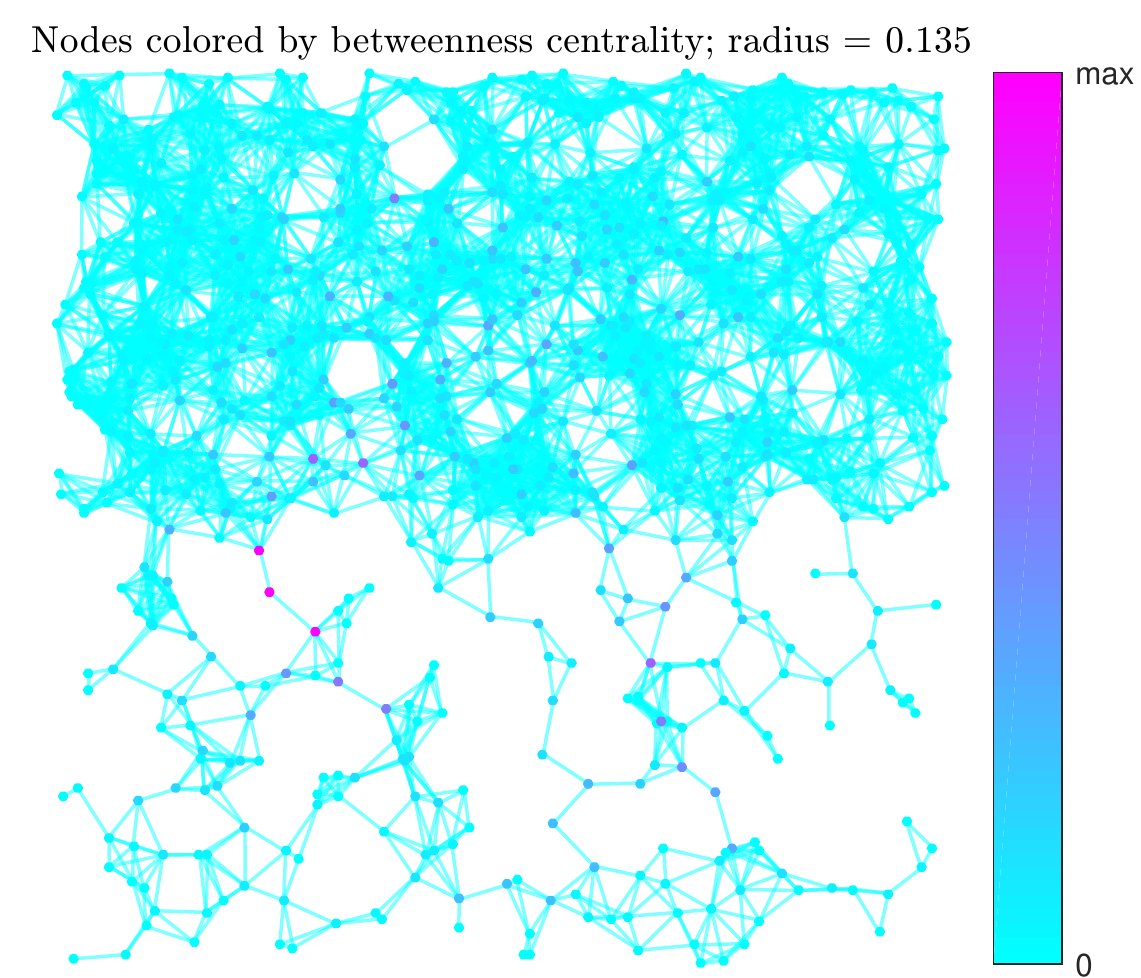}% Here is how to import pix
\includegraphics[trim = 50mm 95mm 50mm 85mm, clip, width=.5\textwidth,keepaspectratio]{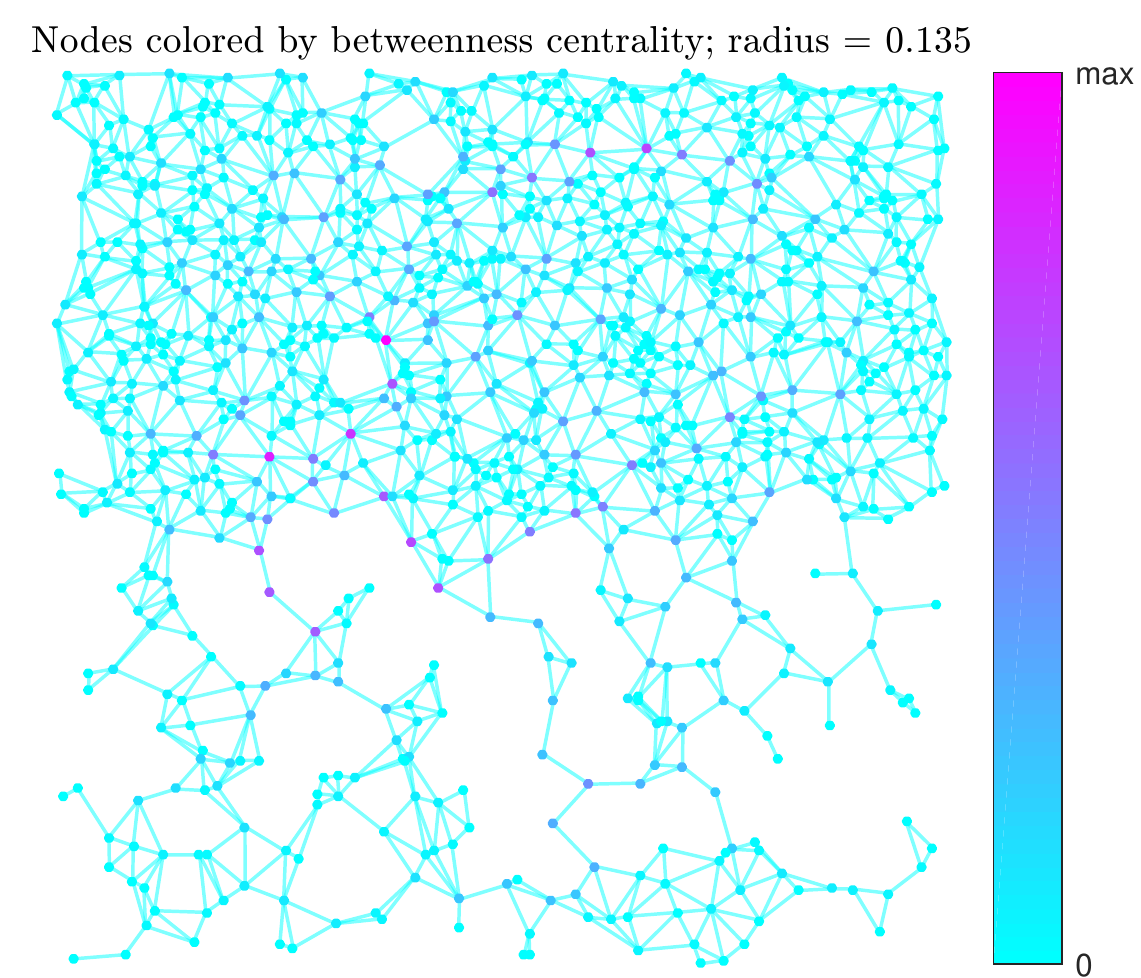}% Here is how to import pix
\caption{ \label{fig:rggrd} (L) $RGG(\xi;r)$ for $\xi$ corresponding to the scenario in Figure \ref{fig:FMM} and $r = 0.135$, slightly above the threshold for connectivity. (R) $RD(\xi;r)$.} 
\end{center}
\end{figure}

Let $\xi_j \in \mathbb{R}^2$ for $1 \le j \le N$, and let $r > 0$. The \emph{random geometric graph} or \emph{disk graph} (RGG; Figure \ref{fig:rggrd}) $RGG(\xi;r)$ has vertices $\xi_j$ and edges $E(RGG(\xi;r)) := \{(\xi_j,\xi_k) : d(\xi_j,\xi_k) \le r\}$ \cite{Haenggi2013,Penrose2003}. By construction, a RGG is both the most effective network topology from the point of view of information exchange, and the least effective network topology from the point of view of infrastructure costs. 

A more conservative topology is based on subgraphs of the \emph{Delaunay graph}. The Delaunay graph $D(\xi)$ has vertices $\xi_j$ and edges defined from a triangulation of the vertices such that no vertex is interior to a circle circumscribed about a triangle \cite{Chen2012,Fuetterling2014,Funke2017}. 
\footnote{
For $\xi_j$ in general position, the Delaunay graph is unique.
}

\begin{figure}[htbp]
\begin{center}
\includegraphics[trim = 50mm 95mm 50mm 85mm, clip, width=.5\textwidth,keepaspectratio]{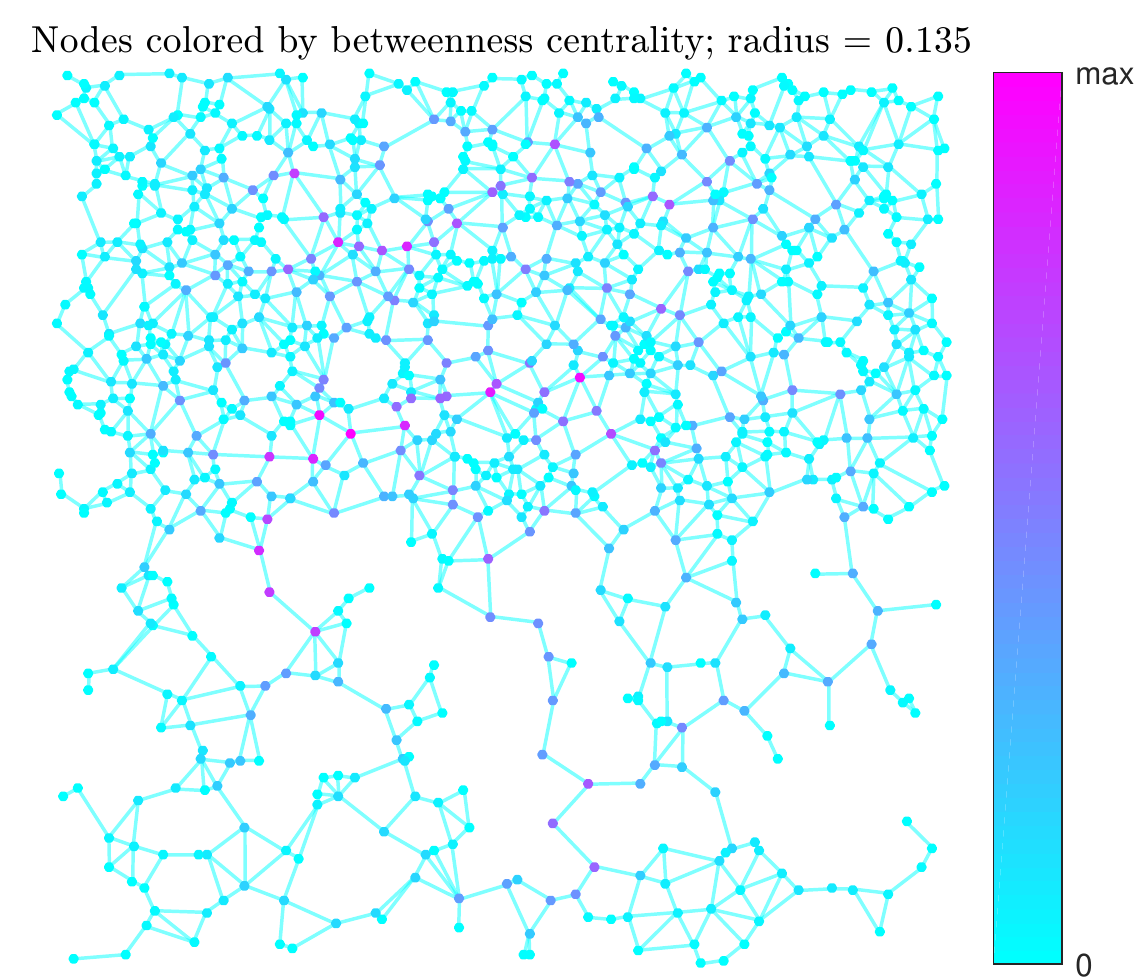}% Here is how to import pix
\includegraphics[trim = 50mm 95mm 50mm 85mm, clip, width=.5\textwidth,keepaspectratio]{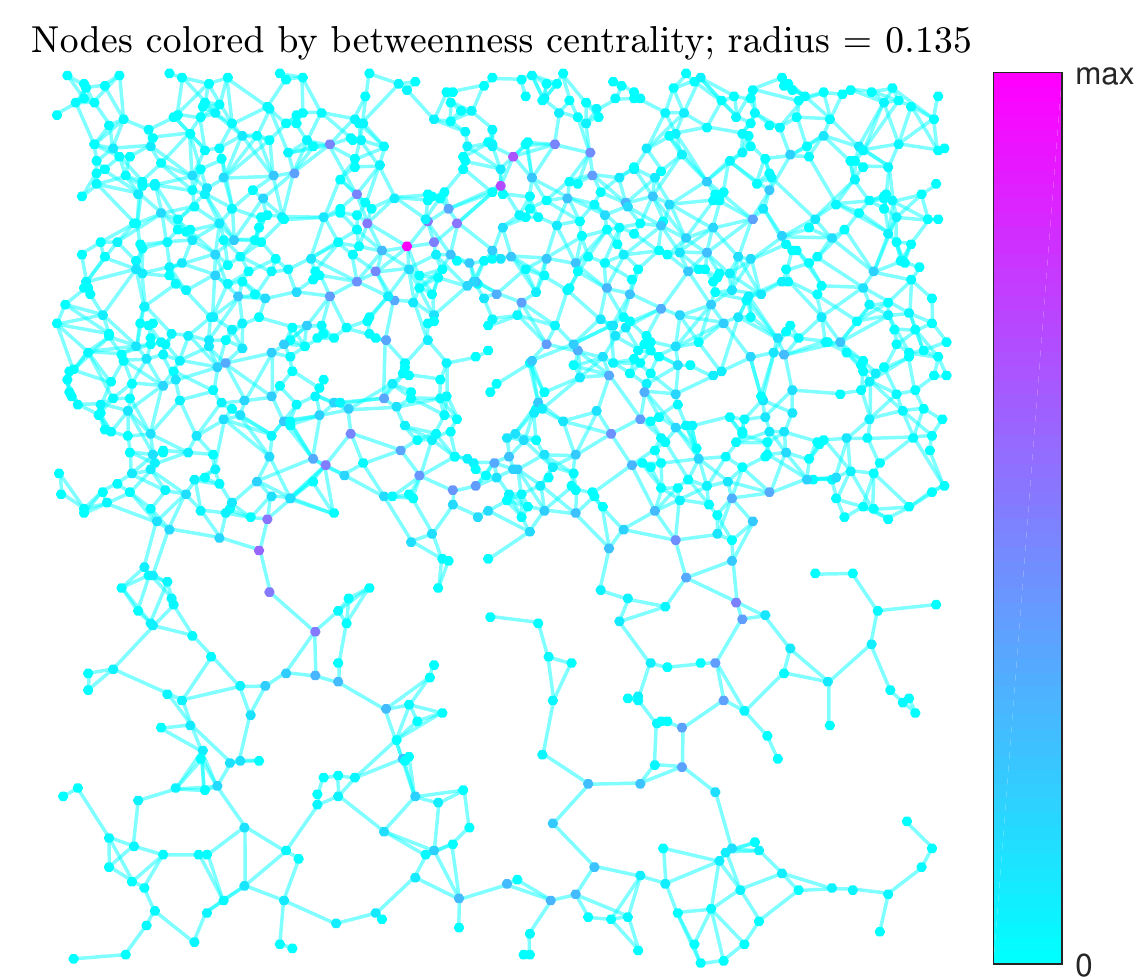}% Here is how to import pix
\caption{ \label{fig:rgrfmn} (L) $RG(\xi;r)$ for $\xi$ corresponding to the scenario in Figure \ref{fig:FMM} and $r = 0.135$, slightly above the threshold for connectivity. (R) $RFMN(\xi;r)$.} 
\end{center}
\end{figure} %

The \emph{Gabriel graph} $G(\xi)$ \cite{Mesbahi2010,Norrenbrock2016} is the unique (for the general position case) subgraph of the Delaunay graph such that each edge corresponds to the diameter of a disk that does not contain any other vertices; it is frequently considered as a potential candidate for ``virtual backbones'' in MANETs. It is worth noting however that $G(\xi)$ and $D(\xi)$ are more computationally expensive to construct than $FMN(\xi)$, and parallelism does not change this.

Because the Delaunay and Gabriel graphs do not have an intrinsic range parameter that will give a granular mechanism for evaluating their performance, we shall focus our attention on the \emph{(minimal) restricted Delaunay graph} (Figure \ref{fig:rggrd}) $RD(\xi;r) := D(\xi) \cap RGG(\xi;r)$ \cite{Avin2008} and the \emph{restricted Gabriel graph} (Figure \ref{fig:rgrfmn}) $RG(\xi;r) := G(\xi) \cap RGG(\xi;r)$. Similarly, we shall consider the \emph{restricted FMN} (Figure \ref{fig:rgrfmn}) obtained along the lines $RFMN(\xi;r) := FMN(\xi) \cap RGG(\xi;r)$. 

The basic evaluation metric we use is the \emph{efficiency} of a graph $G = (V(G),E(G))$, defined as the average inverse distance between distinct vertices, i.e.
\begin{equation}
\label{eq:efficiency}
\text{eff}(G) := \binom{|V(G)|}{2}^{-1} \sum_{\substack{j,k \in V(G) \\ j \ne k}} \frac{1}{d_{jk}},
\end{equation}
where the distance $d_{jk}$ between vertices $j$ and $k$ is computed in the obvious way from a given distance on edges (by default, we may always choose the \emph{hop metric} assigning $1$ to each edge). While the efficiency characterizes how well a network supports information flow \cite{Latora2001}, it neglects costs (e.g., bandwidth, energy, etc.) associated to edges as infrastructure. For this reason we will also consider the \emph{efficiency per edge}, i.e. $\text{eff}(G)/|E(G)|$. Although other normalizations may be more appropriate in certain situations, this particular one strikes a good balance between convenience/generality and detail, especially for the hop metric. 

\begin{figure}[htbp]
\begin{center}
\includegraphics[trim = 20mm 45mm 20mm 40mm, clip, width=\columnwidth,keepaspectratio]{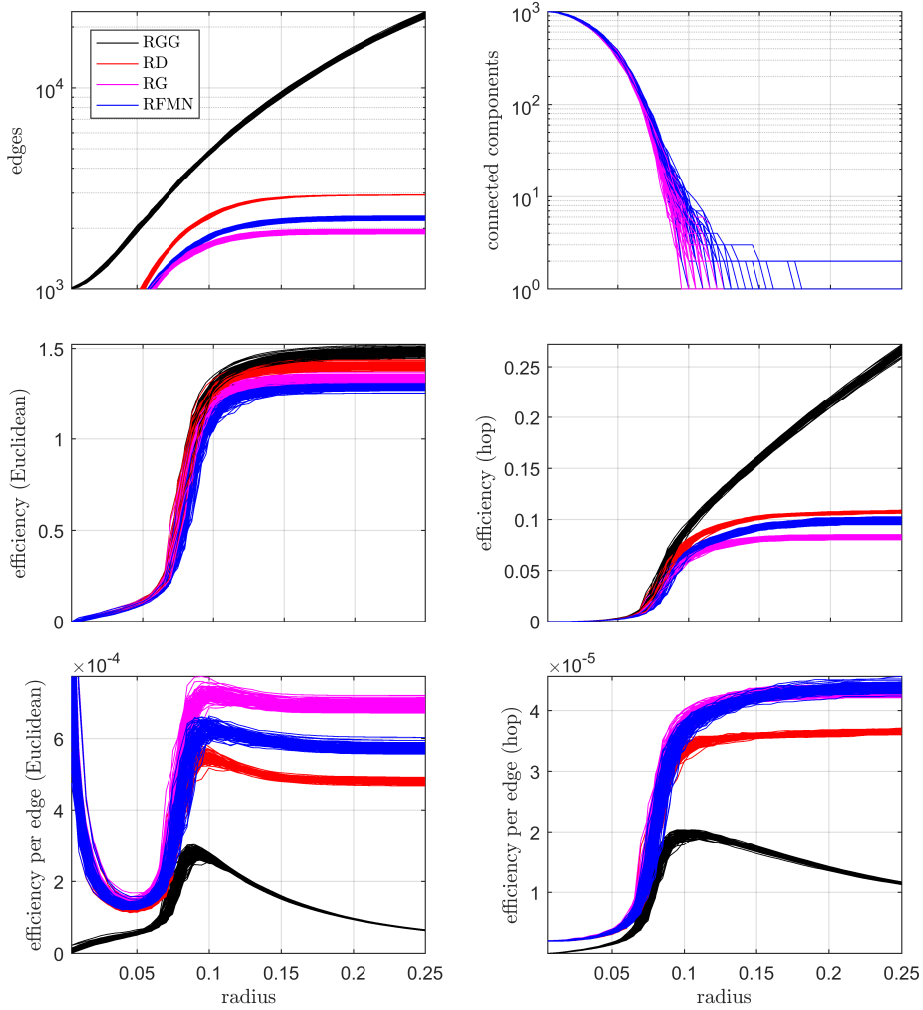}% Here is how to import pix
\caption{ \label{fig:Metrics} Network metrics for $RGG(\xi;r)$ (in black), {\color{red}$RD(\xi;r)$ (in red)}, {\color{magenta}$RG(\xi;r)$ (in magenta)}, and {\color{blue}$RFMN(\xi;r)$ (in blue)} for 100 simulations of $N = 10^3$ uniformly distributed test charges in $[-1,1]^2$. Although $RGG(\xi;r)$ is most efficient, this network performance comes at a high cost in edges, and $RFMN(\xi;r)$ performs well (and for hop efficiency per edge, the best) for all measures of efficiency. Note that $RFMN(\xi;r) = FMN(\xi)$ for sufficiently large $r$ within the range shown. We also have that, e.g. $RD(\xi;r') = D(\xi)$, and though the corresponding $r'$ is outside the range shown, the residual effects are minimal.} 
\end{center}
\end{figure}

Figure \ref{fig:Metrics} shows the metrics above for 100 simulations of $10^3$ uniformly distributed test points in $[-1,1]^2$ subject to the ambient potential from Figure \ref{fig:FMM}. It is apparent from the figure that FMNs and their range-restricted versions are worthy candidates for network backbones in their own right even before accounting for their mobility-specific advantages. Furthermore, although there exist efficient local and parallel algorithms for constructing Delaunay graphs \cite{Chen2012,Fuetterling2014,Funke2017}, their computation and communication complexity and scaling behavior are still inferior to the FMM.

\begin{figure}[htbp]
\begin{center}
\includegraphics[trim = 20mm 45mm 20mm 40mm, clip, width=\columnwidth,keepaspectratio]{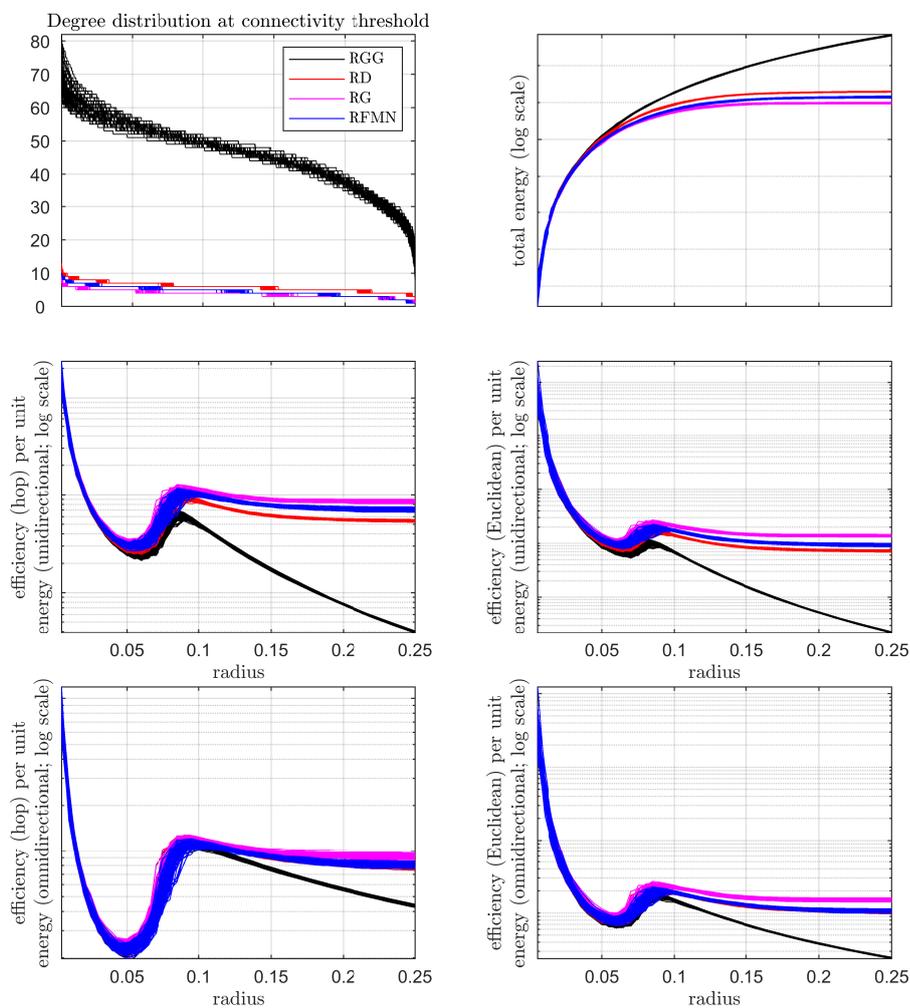}% Here is how to import pix
\caption{ \label{fig:DegreeEnergy} Clockwise from top left, and for the same simulations as Figure \ref{fig:Metrics}: the degree distributions of $RGG(\xi;r)$, $RD(\xi;r)$, $RG(\xi;r)$, and $RFMN(\xi;r)$ for $r$ equal to the connectivity threshold; the total energy (in arbitrary units) required for the networks; the hop efficiency per unit energy, and the Euclidean efficiency per unit energy.} 
\end{center}
\end{figure}

Figure \ref{fig:DegreeEnergy} shows metrics relating to degree distributions and \emph{efficiency per unit energy}, i.e., $\text{eff}(G)/\text{energy}_\bullet(G)$, where (ignoring an irrelevant constant of proportionality) the unidirectional energy for a metric graph $G$ is
\begin{equation}
\label{eq:energyUni}
\text{energy}_\text{uni}(G) := \sum_{(j,k) \in E(G)} d_{jk}^2
\end{equation}
and the omnidirectional energy is
\begin{equation}
\label{eq:energyOmni}
\text{energy}_\text{omni}(G) := \sum_{j \in V(G)} \left( \max_{\substack{k \in V(G) \\ (j,k) \in E(G)}} d_{jk} \right )^2.
\end{equation}
These quantities model the total energy budgets required to transmit uni- and omnidirectional signals, respectively. Figure \ref{fig:DegreeEnergy} highlights that FMNs continue to perform marginally better than Delaunay graphs and marginally worse than Gabriel graphs for energy-normalized measures of network efficiency.

\section{\label{sec:Remarks}Remarks}

By virtue of calculating potentials and forces, the FMM/FMN approach enables dynamic and predictable network topology reconfiguration with minimal cost and effort. In other words, as robots use the FMM to efficiently compute their motion according to a navigation function supplied by a superposition of point charges, the FMN is easily updated and efficiently represented.

Incorporating resilient routing reconfiguration \cite{Wang2010,Decleene2019} on $(F+1)$-connected local subgraphs of the FMN can be done with reasonable computational effort (e.g., the key linear program is quickly and easily solved in MATLAB for realistic networks of $\approx 50$ nodes). This enables virtually instantaneous failover and rerouting in the presence of $\le F$ link failures. Combining this local approach with a separate (perhaps similar) routing protocol to handle wide-area network traffic and obstacle potentials that prevent deterioration of basic connectivity can ensure network integrity and basic quality of service (QoS). These features can render our framework competitive with the approach of \cite{Stephan2017}, which centers on the higher-level functions of network integrity and QoS, and which uses a convex program instead of an algorithmically simpler linear program. Along similar lines, \cite{Zavlanos2007} shows how to construct artificial potentials that discourage loss of connectivity. Although these fields are not harmonic, it is plausible that this idea can be adapted to the present context. 

It is worth pointing out that there are FMM variants for non-harmonic potentials, e.g. power laws, (generalized) multiquadrics \cite{Beatson1997,Ying2006}, or more general kernels \cite{Letourneau2014}, and many of these have actually been applied in the context of interpolation and/or physical simulation. However, using non-harmonic potentials eliminates the automatic guarantee that there are no metastable local minima. We note in particular that the kernel-independent FMM variant of \cite{Ying2004} exploits the existence and uniqueness of solutions to elliptic boundary value problems \cite{Taylor1996} to represent clustered sources in far field based on their behavior on a suitable boundary. This perspective suggests an extension of FMNs to sources modeled by fundamental solutions of elliptic partial differential equations.

\section*{\label{sec:acknowledgements}Acknowledgements}

We thank Brendan Fong, Marco Pravia, and David Spivak for their comments.

%%%%%%%%%%%%%%%%%%%%%%%%%%%%%%%%%%%%%%%%%%%%%%%%%%%%%%%%%%%%%%%%%%%%%%%%%%%%%%%%

%\pagebreak
%
%\section*{Appendix: Springer-Author Discount}
%
%LNCS authors are entitled to a 33.3\% discount off all Springer
%publications. Before placing an order, the author should send an email, 
%giving full details of his or her Springer publication,
%to \url{orders-HD-individuals@springer.com} to obtain a so-called token. This token is a
%number, which must be entered when placing an order via the Internet, in
%order to obtain the discount.
%
%\section{Checklist of Items to be Sent to Volume Editors}
%Here is a checklist of everything the volume editor requires from you:
%
%
%\begin{itemize}
%\settowidth{\leftmargin}{{\Large$\square$}}\advance\leftmargin\labelsep
%\itemsep8pt\relax
%\renewcommand\labelitemi{{\lower1.5pt\hbox{\Large$\square$}}}
%
%\item The final \LaTeX{} source files
%\item A final PDF file
%\item A copyright form, signed by one author on behalf of all of the
%authors of the paper.
%\item A readme giving the name and email address of the
%corresponding author.
%\end{itemize}
\end{document}